\documentclass{elsart}
 \usepackage{natbib}
\usepackage{amsmath}
\usepackage{amssymb}
\usepackage{graphics}
\usepackage{graphicx}
\usepackage{multirow}

\begin{document}

\begin{frontmatter}
\title{Prediction Measures in Beta Regression Models}
\author[1]{Patr\'{\i}cia L.\ Espinheira\corauthref{cor}},
\corauth[cor]{Corresponding author.}
\ead{patespipa@de.ufpe.br}
\author[1]{Luana Cec\'{\i}lia Meireles da Silva},
\author[1]{Alisson de Oliveira Silva}
\address[1]{Departamento de Estat\'{\i}stica, Universidade Federal de Pernambuco,
Cidade Universit\'aria, Recife/PE, 50740--540, Brazil}
\begin{abstract}
We consider the issue of  constructing  PRESS statistics and coefficients of prediction  for a class of beta regression models.
We aim at displaying  measures of predictive power of the model regardless  goodness-of-fit. 
 Monte Carlo simulation results on the finite sample behavior of such measures are provided. 
We also present an application that relates  to the distribution
of natural gas for home usage in S\~ao Paulo, Brazil. Faced with the economic risk of to overestimate or to underestimate the distribution of gas was necessary to construct prediction limits
using beta regression models  \citep{Espinheira2014}. 
Thus, it arises the aim of this work, the selection of best predictive model to construct best prediction limits.
\end{abstract}

\end{frontmatter}

\section{Introduction}
The beta distribution is commonly used to model random variables that assume values in $(0,1)$, such as percentages, rates and proportions. The beta density can display quite different shapes depending on the parameter values. Oftentimes the variable of interest is related to a set of independent (explanatory) variables. \cite{Ferrari2004} introduced a regression model in which the response is beta-distributed, its mean being related to a linear predictor through a link function. The linear predictor includes independent variables and regression parameters. Their model also includes a precision parameter whose reciprocal can be viewed as a dispersion measure. In the standard formulation of the beta regression model it is assumed that the precision is constant across observations. However, in many practical situations this assumption does not hold. \cite{Smithson2006}  consider a beta regression specification in which dispersion is not constant, but is a function of covariates and unknown parameters.  Parameter estimation is carried out by  maximum likelihood (ML) and standard asymptotic hypothesis testing can be easily performed. Practitioners can use the {\tt betareg} package, which is available for the {\tt R} statistical software ({\tt http://www.r-project.org}), for fitting beta regressions.  \cite{Cribari2010} provide an overview of varying dispersion beta regression modeling using the {\tt betareg} package.  

Recently  \cite{Espinheira2014}  built and evaluated bootstrap-based prediction intervals for the class of beta regression models with varying dispersion.  However, a prior approach it is necessary, namely: the selection of 
the model with the best predictive  ability, regardless of the goodness-of-fit. Indeed, the model selection is a crucial step in data analysis, since all inferential performance is based on the selected model.
\cite{Bayer2014a} evaluated the performance of different selection criteria models in samples of finite size in beta regression model, such as Akaike Information Criterion (AIC) \citep{Aka:1973}, Schwarz Bayesian Criterion (SBC) \citep{Schwarz78}, residual sum of squares (RSS), and  various functions of RSS  such as  the coefficient of determination, $R^2$ and the  adjusted $R^2$.
However, these methods  do not offer any insight about the quality of the predictive values.
In this context, \cite{All:1974}, proposed the PRESS (Predictive Residual Sum of Squares) criterion, that can be used as an indication of the predictive power of a model.  The PRESS statistic  is independent from the goodness-of-fit  of the model, since that its calculation is made by leaving out  the observations  that the model  is trying to predict.  The PRESS statistics can be viewed as a sum of squares of external residuals. Thus, similarly of the approach of $R^2$  \cite{Mediavilla2008}  proposed a coefficient of prediction based on PRESS namely $P^2$.  The  $P^2$ statistic can be used to select models from a predictive perspective adding important information about the predictive ability of the model in various scenarios.  
\section{On beta regression residuals}
Let $y_1, \ldots, y_n$ be independent random variables such that each $y_t$, for $t=1,\ldots,n$, is beta distributed,
i.e., each $y_t$ has density function given by
\begin{equation}\label{eq1}
f(y_t; \mu_t, \phi_t) = \frac{\Gamma(\phi_t)}{\Gamma(\mu_t\phi_t) 
\Gamma((1-\mu_t)\phi_t)} y_t^{\mu_t\phi_t-1}(1-y_t)^{(1-\mu_t)\phi_t-1},\quad
0 < y_t < 1,
\end{equation}
where $0 < \mu_t < 1$ and $\phi_t > 0$.
Here, ${\rm E}(y_t) = \mu_t$ and ${\rm Var}(y_t) = {V(\mu_t) / (1+\phi_t)}$, where $V(\mu_t) = \mu_t(1-\mu_t)$.
In the beta regression model introduced by \cite{Ferrari2004} the mean of $y_t$ can be written as 
\begin{equation}\label{eq2}
g(\mu_t) =x_t^{\!\top}\beta = \eta_t. 
\end{equation}
In addition to the relation given in \eqref{eq2}, it is possible to assume that the precision parameter is not constant and write 
\begin{equation}\label{eq3}
h(\phi_t)  = z_t^{\!\top}\gamma = \vartheta_t.
\end{equation}
In \eqref{eq2} and \eqref{eq3}, $\eta_t$ and $\vartheta_t$ are linear predictors,
$\beta = (\beta_1, \ldots, \beta_k)^{\!\top}$ and $\gamma = (\gamma_1, \ldots, \gamma_q)^{\!\top}$ are unknown parameter vectors
($\beta \in \mathbb{R}^k$; $\gamma \in \mathbb{R}^q$),  
$x_{t1}, \ldots, x_{tk}$ and $z_{t1}, \ldots, z_{tq}$ are fixed covariates ($k+q < n$) and $g(\cdot)$ and $h(\cdot)$ are link functions,
which are strictly increasing and twice-differentiable.  

The PRESS statistic is based on sum of external residuals obtained from  exclusion of observations.
 For beta regression models \cite{Ferrari2011} present a standardized residual obtained
using Fisher's scoring iterative algorithm for $\beta$ under varying dispersion. Here, we propose a new residual  based on a combination of ordinary residuals obtained
using the algorithms for $\beta$ and $\gamma$ under varying dispersion.
At the outset, consider the Fisher's scoring iterative algorithm for estimating $\beta$ (see the Appendix~\ref{apendice}). From \eqref{A5} it follows that the $m$th step of the scoring scheme is  
\begin{equation}\label{eq4}
\beta^{(m+1)} =\beta^{(m)} +  (X^{\!\top}\Phi^{(m)}W^{(m)}X)^{-1}\Phi^{(m)} X^{\!\top}T^{(m)} (y^*-{\mu^*}^{(m)}),
\end{equation}
where the $t$th elements of the vectors $y^*$ and $\mu^*$ are given, respectively, by 
\begin{equation}\label{eq5}
y_t^* = \log\{ y_t / (1-y_t)\} \,\, \,\, {\rm and}\,\,\,\, \mu_t^* = \psi(\mu_t\phi_t)- \psi((1-\mu_t)\phi_t),
\end{equation}
\noindent $\psi(\cdot)$ denoting the digamma function, i.e., $\psi(u) =
{\rm d} \log \Gamma(u)/ {\rm d}u$ for $u> 0$. The matrices $T$ and $W$ are 
given in \eqref{A1} and \eqref{A3}, respectively, $X$ is an $n\times k$ matrix whose 
$t$th row is $x_t^{\!\top}$  and $\Phi = {\rm diag}(\phi_1,\ldots,\phi_n)$. Note that ${\mu}^*_t = {\rm E}(y^*_t)$ (see \eqref{A6}; Appendix~\ref{apendice}). Similarly, from \eqref{A5} it follows that the $m$th step of the scoring scheme for $\gamma$ is given by
\begin{equation}\label{eq6}
\gamma^{(m+1)} =\gamma^{(m)} +  (Z^{\!\top}D^{(m)}Z)^{-1} Z^{\!\top}H^{(m)}a^{(m)},
\end{equation}
where the $t$th element of $a_t$ is give by
\begin{equation}\label{eq7}
a_t = \mu_t(y_t^{*} - \mu_{t}^{*}) + \log(1 - y_t) - \psi((1 - \mu_t)\phi_t) + \psi(\phi_t)
\end{equation}
and  the matrices  $H$ and  $D$ are 
given in \eqref{A2} and \eqref{A4}, respectively, and $Z$ is an $n\times q$ matrix that
$t$th row is $z_t^{\!\top}$. 
It is possible to write the iterative schemes in \eqref{eq4} and \eqref{eq6} in terms of weighted least squares 
regressions, respectivelly as 
$\beta^{(m+1)} =(X^{\!\top}\Phi^{(m)}W^{(m)}X)^{-1}\Phi^{(m)} X^{\!\top}W^{(m)}u_1^{(m)}$ and 
$\gamma^{(m + 1)} = (Z^{\top}D^{(m)}Z)^{-1}Z^{\top}D^{(m)}u_2^{(m)}$.
Where
$u_1^{(m)} =\eta^{(m)} + {W^{-1}}^{(m)}T^{(m)}(y^*-{\mu^*}^{(m)})$, with
$\eta = (\eta_1,\ldots,\eta_n)^{\!\top} = X\beta$, $u_2^{(m)} =\vartheta^{(m)} + {D^{-1}}^{(m)}H^{(m)}a^{(m)}$, with $\vartheta = (\vartheta_1,\ldots,\vartheta_n)^{\!\top}=$ $ Z\gamma$ and  $a_t$ given in \eqref{eq7}.
Upon convergence, 
\begin{equation}\label{eq8}
    \begin{split}
&\,\widehat{\!\beta} =(X^{\!\top}\,\widehat{\! \Phi}\,\widehat{\! W}X)^{-1} \,\widehat{\! \Phi}X^{\!\top}\,\widehat{\! W}u_1 \quad \mbox{and} \quad \,\widehat{\!\gamma} =(Z^{\!\top}\,\widehat{\! D}Z)^{-1} Z^{\!\top}\,\widehat{\! D}u_2, \quad \text{where}\\
& u_1={\,\widehat{\! \eta}} + {\,\widehat{\! W}}^{-1}{\,\widehat{\! T}}(y^*-{\widehat{{\mu}}^*})\quad \mbox{and} \quad u_2={\,\widehat{\! \vartheta}} + {\,\widehat{\! D}}^{-1}{\,\widehat{\! H}}\widehat{a}.
\end{split} 
\end{equation}
Here, $\widehat{W}$, $\widehat{T}$, $\widehat{H}$ and $\widehat{D}$ are the matrices $W$, $T$, $H$ and $D$ respectively, evaluated at the maximum likelihood estimator. 
We note that $\,\widehat{\!\beta}$ and $\,\widehat{\!\gamma}$ in \eqref{eq8} can be viewed as the least squares estimates of $\beta$ and $\gamma$ obtained by regressing $\,\widehat{\Phi}^{1/2} \,\widehat{\! W}^{1/2}u_1$ and ${\,\widehat{\! D}}^{1/2}u_2$ on $\,\widehat{\Phi}^{1/2}\,\widehat{\! W}^{1/2}X$ and ${\,\widehat{\! D}}^{1/2}Z$, respectively. The residuals ordinary obtained of interactive process of  $\beta$ and $\gamma$ are given by
$r^{\beta} =\widehat{\Phi}^{1/2}{\,\widehat{\! W}}^{1/2}(u_1 - {\,\widehat{\! \eta}}) ={\,\widehat{\Phi}}^{1/2} {\,\widehat{\! W}}^{-1/2}{\,\widehat{\! T}}(y^*-{\widehat{{\mu}}^*})$ and $r^{\gamma} ={\,\widehat{\! D}}^{1/2}(u_2 - {\,\widehat{\! \vartheta}}) = {\,\widehat{\! D}}^{-1/2}{\,\widehat{\! H}}\widehat{a}$, respectively. Hence, using the definitions of the matrices given from \eqref{A1} to \eqref{A5}, we can rewrite the residuals obtained from the iterative process of $\beta$ and $\gamma$ respectively, as
\begin{equation}\label{eq9}
r^{\beta}_t = {\frac{y^*_t - {\widehat{{\mu}}^*}_t} { \sqrt{\widehat{v}_t }}}\quad \mbox{and} \quad r^{\gamma}_t = \frac{\widehat a_t}{\sqrt{\widehat{\varsigma}}_t}, 
\end{equation}
where $v_t$ and $\varsigma_t$ are given in \eqref{A3} and \eqref{A4}, respectively. 
Thus, we propose a new residual based on $r^{\beta}$ and $r^{\gamma}$, which we shall refer to as the combined residual
$r^{\beta\gamma}_t = (y^{*}_t - \hat{\mu}^{*}_t) + \widehat a_t $
\noindent where $y_t^*$ and $\mu_t^*$ are given in \eqref{eq5}.  Assuming that  $\mu_t$ and $\phi_t$ are known and from \eqref{A6} to \eqref{A10} it follows that   ${\rm Var} ({r^{\beta\gamma}_t}) = \zeta_t$, with 
\begin{equation}\label{eq10}
\zeta_t = (1  + \mu_t)^2  \psi^{\prime}(\mu_t\phi_t) + \mu_t^2
\psi^{\prime}((1 - \mu_t)\phi_t) - \psi^{\prime}(\phi_t).  \end{equation} Then, we can define the following standardized combined residual:
\begin{equation}\label{eq11}
r^{\beta\gamma}_{p,t} = \frac{(y^{*}_t - {\,\widehat{\! \mu}}^{*}_t) +{\,\widehat{\! a}}_t}{\sqrt{\,\widehat{\! \zeta_t}}}
\end{equation}
Here, $\widehat{\zeta_t}$ is $\zeta_t$ in \eqref{eq10} evaluated at  ${\,\widehat{\! \mu}}_t$ e ${\,\widehat{\! \phi}}_t$.  It is important to note that when $\phi$ is constant it is only necessary replace $\phi_t$ by $\phi$ at all elements of \eqref{eq11}. 
We should emphasize that here we are just interested in evaluating the $r^{\beta\gamma}_p$ in the composition of the PRESS statistic.  
\section{$P^2$ Statistics }\label{sec3}
Consider the linear model,
$Y = X\beta + \varepsilon$
\noindent where $Y$ is a vector $n \times 1$ of  responses, $X$ is a known matrix of covariates of dimension $n \times p$, $\beta$ is the parameter vector of dimension $p \times 1$  and $\varepsilon$ is  a vector  $n \times 1$ of errors   distributed as $N_n(0; \sigma^2I_n)$. Let $\widehat{\beta} = (X^{\top}X)^{-1}X^{\top}y$, $e_t = y_t - x_t^{\top}\hat{\beta}$, $\widehat{y} = x_t^{\top}\widehat{\beta}$ and let $\hat{\beta}_{(t)}$ be the estimate of $\beta$ without the i$th$ observation and $\widehat{y}_{(t)} = x_t^{\top}\widehat{\beta}_{(t)}$ be the case deleted predicted value of the response when the independent variable has value $x_i$. Thus, for multiple regression
$PRESS = \sum_{t = 1}^{n}(y_t - \widehat{y}_{(t)})^2$ which can be rewritten as $PRESS = \sum_{t = 1}^{n}(y_t - \widehat{y}_t)^2/(1 - h_{tt})^2$, where $h_{tt}$ is the t$th$ diagonal element of the matrix $X(X^{\top}X)^{-1}X^{\top}$. 

In the beta regression model $\widehat \beta$ in \eqref{eq8} can be viewed as the least squares estimate of  $\beta$ obtained
by regressing 
\begin{equation}\label{eq12}
\check{y} =\widehat\Phi^{1/2} {\,\widehat{\! W}}^{1/2}u_1\,\,\,  \text {on} \,\,\check{X} = \widehat\Phi_t^{1/2}{\,\widehat{\! W}}^{1/2}X.  
\end{equation}
Thus, the prediction error is
$\check{y}_t - \widehat{\check{y}}_{(t)} = \widehat\phi_t^{1/2}\widehat{w}^{1/2}_t u_{1,t} - \widehat\phi_t^{1/2}\widehat{w}^{1/2}_t x_t^{\top}\widehat{\beta}_{(t)}$. Using the ideas proposed by \citep{pregibon1981} and  fact that 
 $$\widehat{\beta}_{(t)} = \widehat{\beta} - \frac{{(X^{\top}\widehat \Phi{\,\widehat{\! W}}X)^{-1}x_t \widehat\phi_t^{1/2}{\,\widehat{\! w}}_t^{1/2} r^{\beta}_t}}{({1 - h_{tt}^*})},$$  
 where $r_t^{\beta}$ is given in \eqref{eq9} and $h^*_{tt}$ is the $t$th diagonal element of  $$H^* =
 ({\,\widehat{\! W}}{\,\widehat{\! \Phi}})^{1/2}X(X{\,\widehat{\! \Phi}}{\,\widehat{\! W}}X)^{-1}X^{\top}({\,\widehat{\! \Phi}}{\,\widehat{\! W}})^{1/2}$$
it then follows that
$\check{y}_t - \hat{\check{y}}_{(t)} = \{{r^{\beta}_t}\}/({1 - h_{tt}^*}).
$
Finally, for the beta regression model the PRESS statistic is given by
\begin{equation}\label{eq13}
PRESS = \sum^{n}_{t=1}(\check{y}_t - \hat{\check{y}}_{(t)})^2 = \sum^{n}_{t=1}\left(\frac{r^{\beta}_t}{1 - h_{tt}^*}\right)^2.
\end{equation}
In (\ref{eq13}) the  $t$th observation is not used in fitting the regression model to predict $y_t$, then both the external predicted values $\hat{y}_{(t)}$ and the external residuals $e_{(t)}$ are independent of $y_t$. This fact  enables the PRESS statistic to be a true assessment of the prediction capabilities of the regression model regardless of the overall quality of the fit of the model.

Considering the same approach of the  coefficient of determination $R^2$, we can think in a prediction coefficient based on PRESS, namely 
\begin{equation}\label{eq14}
P^2 = 1 - \frac{PRESS}{SST_{(t)}},
\end{equation}
\noindent wherein $SST_{(t)} = \sum_{t=1}^n(y_t- \bar{y}_{(t)})^2$ and  $\bar{y}_{(t)}$ is the arithmetic average of the ${y}_{(t)},\,t = 1,\ldots, n$.  It can be shown that $SST_{(t)} = (n/n-p)^2 SST$, wherein $p$   is the number of model parameters.
In the beta regression model with varying dispersion,  
$SST = \sum_{t=1}^n(\check{y}_t- \bar{\check{y}})^2$, $\bar{\check{y}}$ is the  is the arithmetic average of the  $\check{y}_t = \widehat{\phi}_t^{1/2}\widehat{w}_t^{1/2} u_{1,t},\,t = 1,\ldots, n$ given in \eqref{eq12}  and $p = k + q$.

\cite{CookWeisberg82book} suggest  other versions of PRESS statistics based on different residuals. Thus, we  present another version of PRESS statistics and   $P^2$ associated considering a new residual presented in \eqref{eq11}, such that 
\begin{equation}\label{eq15}
PRESS_{\beta\gamma}  = \sum^{n}_{t=1}\left(\frac{r^{\beta\gamma }_{p,t}}{1 - h_{tt}^*}\right)^2\quad {\rm and} \quad
P^2_{\beta\gamma} = 1 - \frac{PRESS_{\beta\gamma}}{SST_{(t)}}, 
\end{equation}
\noindent respectively. 
It is noteworthy that the measures $R^2$ and $P^2$ are distinct, since that the $R^2$ propose to measure the quality of fit of the model and the $P^2$ and $P^2_{\beta\gamma}$ measure the predictive power. Additionally, $P^2$ and $P^2_{\beta\gamma}$ are not positive measure. In fact, the $\text {PRESS/SST}_{(t)}$  is  a positive quantity, thus the  $P^2$ and the $P^2_{\beta\gamma}$ associated given in \eqref{eq14} and  \eqref{eq15}, respectively,  take
values in $(-\infty; 1]$. The closer to one the better is the predictive power of the model.
In order to
check the goodness-of-fit of the estimated model, we used the approach suggested by \cite{Bayer2014a} for beta regression models with varying dispersion, a
version of  $R^2$ based on likelihood ratio, given by:
$
R^2_{LR} = 1 - ({L_{null}}/{L_{fit}})^{2/n},
$
 wherein $L_{null}$ is the maximum likelihood achievable (saturated model) and $L_{fit}$ is the achieved by the model under
investigation. 
\subsection{Monte Carlo results}
The Monte Carlo experiments were carried out using using both fixed and varying dispersion beta regressions as data generating processes. All results are based on 10,000 Monte Carlo replications. Table~\ref{T:T1} contains numerical results for the fixed dispersion beta regression model as data generating processe, given
by
$$
\log \left(\frac{\mu_t}{1-\mu_t}\right) = \beta_1 + \beta_2\,x_{t2} + \beta_3\,x_{t3} + \beta_4\,x_{t4} + \beta_5\,x_{t5}, \quad t = 1, \ldots, n ,
$$
The covariate values were independently obtained as
random draws of the following distributions: $X_{ti} \sim U(0,1)$, $i = 2, \ldots, 5$  and
were kept fixed throughout the experiment.
 The precisions, the sample sizes and  the mean  response are, respectively,  $\phi = (50, 148, 400)$, $n = (40, 80, 120)$, $\mu \in (0.005, 0.12)$, $\mu \in (0.90, 0.99)$ and $\mu\in (0.20, 0.88)$.
To investigate the performances of  statistics in the omission of covariates, we considered the Scenarios 1, 2 and 3, in which are omitted, three, two and one covariate, respectively. In the fourth scenario the estimated model is correctly specified.
Additionally we calculate the $R^2_{LR}$  for the same scenarios.
The results in Table~\ref{T:T1} show that the values of all statistics increase as important covariates are included in the model.
Statistics behave similarly as the sample size and the precisions values indicating that the most important factor is the correct specification of the model.   Considering the three ranges for the $\mu$ it should be noted that the statistic values are considerably larger when $\mu \in (0.20,0.88)$  and  the values approaching one when the estimated model is closest to the true model. For instance, in Scenario 4 for $n = 40$, $\phi = (50, 148, 400)$ the values of $P^{2}$  and  $R^2_{LR}$  are, respectively,  (0.8354, 0.9357,  0.9748)  and   (0.8349, 0.9376, 0.9758). 

The statistics finite sample behavior substantially change when $\mu \in (0.90; 0.99)$. It is  noteworthy 
the reduction of the statistic values, revealing the difficulty  in to fit the model and make prediction when  $\mu \approx 1$.  
Indeed, in this range of $\mu$ is more difficult to make prediction that to fit the model. For example, in Scenario 1, when three covariates are omitted from the model,
when $n = 40$  and $\phi = (50, 148, 400)$  the $P^{2}$ values equals, 0.0580, 0.0636 and 0.0972  whereas the 
the $R^2_{LR}$  values are 0.1553, 0.1999 and 0.2496, respectively. Similar results were obtained for   $n = 80, 120$.  Even when for the correctly specified four covariate
model (Scenario 4) the predictive power of the model is more affected than the quality of fit of the model by the fact of $\mu \approx 1$.
In this situation, it is  noteworthy that the finite
sample performances predictive power model improve when the value of the precision parameter increases.
For instance, when  $n=120$  and  $\phi = (50, 148, 400)$  we have $P^{2} = (0.0272, 0.2222,  0.5622)$ and
$P^{2}_{\beta\gamma} = (0.063, 0.5348,  0.8381)$, respectively.    
Here it is possible see that the   $P^{2}_{\beta\gamma}$  statistic always shows larger values than the
 $P^{2}$  statistic when the mean
responses are close to of the upper limit of the standard unit interval.  However, the two measures behave similarly
when used to investigate model misspecification.

The same difficulty in obtaining predictions and in fitting the regression model  occurs when $\mu \in (0.005, 0.12)$. Once again the greatest difficulty   lies on the predictive power of the model.
It is also noteworthy  that when  $\mu \approx 0$ the point prediction becomes even less reliable than
when $\mu \approx 1$, since the  $P^2$ and $P^2_{\beta\gamma}$ values decreased substantially and become considerably distant from the $R^2_{LR}$  values.
When the mean
responses are close to of the lower limit of the standard unit interval, the  $P^2_{\beta\gamma}$ seems to be more able in identify  poor predictions.    For instance, in Scenario 4 (model correctly specified; four covariates)
when $n=120$ and $\phi = (50, 148, 400)$, we have  $P^{2} = (0.0464,0.2716, 0.6322)$ and
$P^{2}_{\beta\gamma}= (0.0362, 0.0468, 0.0603)$, respectively.

\begin{table}
\caption{\label{T:T1} Statistic values. True model:
$g(\mu_t) =\log( {\mu_t}/{(1-\mu_t)}) = \beta_1 + \beta_2\,x_{t2} + \beta_3\,x_{t3} + \beta_4\,x_{t4} + \beta_5\,x_{t5},$ $\,t=1,\ldots,n$, $\phi$ fixed. Misspecification: omitted covariates (Scenarios 1, 2 and 3).}
  {\tiny
 \begin{tabular}{*{14}{|c|}} 
&\text{Scenarios}&\multicolumn{3}{|c}{\text{Scenario 1}}&\multicolumn{3}{|c}{\text{Scenario 2}}&\multicolumn{3}{|c}{\text{Scenario 3}}&\multicolumn{3}{|c}{\text {Scenario 4}}\\\hline
&Estimated   &\multicolumn{3}{|c}{$g(\mu_t) = \beta_1 + \beta_2\,x_{t2}$}&\multicolumn{3}{|c}{$g(\mu_t) = \beta_1 + \beta_2\,x_{t2}$}&\multicolumn{3}{|c}{$g(\mu_t) = \beta_1 + \beta_2\,x_{t2}$}&\multicolumn{3}{|c}{$\hskip-0.1trueing(\mu_t) = \beta_1 + \beta_2\,x_{t2}+\hskip-0.1truein$}\\
&model&\multicolumn{3}{|c}{}&\multicolumn{3}{|c}{$+ \beta_3\,x_{t3}$}&\multicolumn{3}{|c}{$+ \beta_3\,x_{t3} + \beta_4\,x_{t4}$}&\multicolumn{3}{|c}{$\hskip-0.05truein\beta_3\,x_{t3} + \beta_4\,x_{t4}+\beta_5\,x_{t5}\hskip-0.1truein$}\\\hline
$\mu$& \multicolumn{13}{|c}{$\mu \in(0.20, 0.88)$}\\ \hline
$n$&$\phi$&50&150&400&50&150&400&50&150&400&50&150&400\\ \hline 
40 &$P^2$        &0.359&0.392&0.406&0.457&0.501&0.518&0.595&0.655&0.679&0.835&0.935&0.974\\
  &$P^2_{\beta\gamma}$      &0.454&0.471&0.478&0.567&0.599&0.611&0.704&0.754&0.774&0.856&0.938&0.974\\
  &$R^2_{LR}$ &0.354&0.390&0.405&0.467&0.514&0.532&0.613&0.674&0.697&0.857&0.946&0.979\\\hline
80 &$P^2$        &0.341&0.377&0.392&0.439&0.487&0.505&0.575&0.642&0.668&0.819&0.929&0.972\\
  &$P^2_{\beta\gamma}$      &0.437&0.457&0.465&0.551&0.587&0.601&0.689&0.745&0.768&0.842&0.932&0.971\\
  &$R^2_{LR}$&0.351&0.389&0.404&0.462&0.512&0.531&0.605&0.671&0.696&0.848&0.942&0.977\\\hline
120&$P^2$        &0.335&0.372&0.387&0.432&0.482&0.501&0.569&0.638&0.664&0.813&0.927&0.971\\
  &$P^2_{\beta\gamma}$      &0.431&0.452&0.460&0.546&0.583&0.598&0.685&0.742&0.765&0.838&0.930&0.970\\
  &$R^2_{LR}$&0.350&0.389&0.404&0.460&0.511&0.531&0.603&0.670&0.696&0.845&0.941&0.977\\\hline
$\mu$& \multicolumn{13}{|c}{$\mu \in (0.90, 0.99) $}\\ \hline
$n$&$\phi$&50&150&400&50&150&400&50&150&400&50&150&400\\ \hline 
40 &$P^2$        &0.058&0.063&0.097&0.062&0.070&0.117&0.065&0.205&0.409&0.071&0.296&0.610     \\
  &$P^2_{\beta\gamma}$      &0.092&0.112&0.217&0.106&0.152&0.298&0.109&0.445&0.711&0.132&0.601&0.858\\
  &$R^2_{LR}$&0.155&0.199&0.249&0.225&0.292&0.364&0.350&0.486&0.621&0.441&0.619&0.794\\\hline
80 &$P^2$        &0.033&0.037&0.072&0.038&0.044&0.097&0.035&0.165&0.385&0.037&0.240&0.574\\
  &$P^2_{\beta\gamma}$      &0.067&0.080&0.192&0.081&0.115&0.277&0.069&0.404&0.699&0.079&0.551&0.843\\
  &$R^2_{LR}$&0.149&0.195&0.246&0.212&0.283&0.358&0.329&0.471&0.612&0.412&0.597&0.781\\\hline
120&$P^2$        &0.025&0.028&0.063&0.030&0.036&0.090&0.025&0.151&0.376&0.027&0.222&0.562\\
  &$P^2_{\beta\gamma}$      &0.058&0.069&0.184&0.072&0.103&0.270&0.057&0.390&0.694&0.063&0.534&0.838\\
  &$R^2_{LR}$&0.147&0.194&0.245&0.207&0.280&0.357&0.322&0.466&0.609&0.403&0.591&0.777\\\hline
$\mu$      & \multicolumn{13}{|c}{$\mu \in(0.005, 0.12)$}\\ \hline
$n$&$\phi$&50&150&400&50&150&400&50&150&400&50&150&400\\ \hline 
40 &$P^2$        &0.067&0.055&0.080&0.072&0.048&0.070&0.072&0.144&0.285&0.079&0.327&0.663\\
  &$P^2_{\beta\gamma}$      &0.044&0.043&0.044&0.049&0.041&0.035&0.061&0.067&0.073&0.076&0.093&0.111\\
  &$R^2_{LR}$&0.214&0.252&0.294&0.274&0.327&0.381&0.378&0.482&0.576&0.526&0.700&0.847\\\hline
80 &$P^2$        &0.044&0.031&0.057&0.050&0.028&0.057&0.046&0.113&0.269&0.046&0.271&0.632\\
  &$P^2_{\beta\gamma}$      &0.022&0.021&0.022&0.025&0.020&0.017&0.029&0.037&0.047&0.036&0.046&0.060\\
  &$R^2_{LR}$&0.209&0.249&0.292&0.263&0.320&0.377&0.361&0.470&0.568&0.504&0.683&0.838\\\hline
120&$P^2$        &0.037&0.023&0.049&0.044&0.022&0.053&0.037&0.101&0.262&0.036&0.252&0.621\\
  &$P^2_{\beta\gamma}$      &0.015&0.014&0.015&0.018&0.013&0.011&0.019&0.027&0.038&0.023&0.032&0.043\\
  &$R^2_{LR}$&0.207&0.248&0.291&0.259&0.317&0.375&0.356&0.465&0.566&0.497&0.677&0.834\\\hline
\end{tabular}}
\end{table}

We have also carried out Monte Carlo simulations using a varying dispersion
beta regression model, in which we increased the number
of covariates, used different covariates in the mean and precision submodels.
In this case the data generating process and the postulated model is the same .
We report
results for $\lambda = (20, 50, 100)$,  $n = (40, 80, 120)$,  $\mu \in (0.20, 0.88)$,  $\mu \in (0.90, 0.99)$ and $\mu \in (0.005, 0.12)$. Here, 
\begin{equation}\label{eq16}
\lambda =  \frac{\phi_{\max}}{\phi_{\min}}={\max\limits_{t=1,\ldots,n}\{\phi_t \}\over\min\limits_{t=1,\ldots,n}\{\phi_t\}},
\end{equation}
is the measure the intensity of nonconstant dispersion. The covariate values in the mean submodel and in the precision submodel were obtained as random draws
from the ${\cal U}(0,1)$  and ${\cal U}(-0.5,0.5)$ distributions, respectively,  such that
the covariate values in the two submodels are not the same.
At the end, we also considered a covariate values generated from $t_{(3)}$ (Student's t-distribution with 3 degrees of freedom). The results are presented in Table~\ref{T:T2}.
We should emphasize that  were generated  only $n=40$ covariates values and the  $n=80, 120$ covariates values
are replications of original set. In this sense, the intensity of nonconstant dispersion remains  the same over the sample size.

When the mean
responses are scattered on the standard unit interval ($\mu \in (0.20, 0.88)$) the three statistics display similar values.
It seems that neither the degree of  intensity of nonconstant dispersion nor the simultaneous increase in the number of covariates in the two submodels
noticeably affect the predictive power and fit of the model when the  sample size is fixed.
However, it is noteworthy a reduction of statistic values  when  the response
values are close to one or close to zero, making clear the difficulty in fitting the regression model and obtaining good predictions when  $\mu \approx 1$  or $\mu \approx 1$ and the precision is modelled. 
The minor values of $P^2_{\beta\gamma}$ statistic reveals the problem in to make good predictions when  $\mu \approx 0$, whereas when    
$\mu \approx 1$ this problem is singled out by smaller values of $P^2$ statistic. Here, the model fit is more affect when the
number of covariates increases simultanealy in the two submodels.
For instance consider $n=40$, $\lambda = 100$ and $\mu \in (0.005, 0.12)$. 
At the Scenario 5 (one covariate in both submodels),  we have $P^2$ = 0.8117, $P^2_{\beta\gamma}$ = 0.3677  and $R^2_{LR}$= 0.8228.  
Whereas in Scenario 8 (four covariate in both submodels) we have $P^2$ = 0.8627, $P^2_{\beta\gamma}$ = 0.4863  and $R^2_{LR}$= 0.6447;

We also displayed in Table~\ref{T:T2}  the statistic values when the model is correctly specified, but we  introduced leverage points in the data. To that end, only  the  $X_2$ values were obtained as
random draws of the  $t_{(3)}$  distribution  and concerned ourselves  with $\mu \in (0.20, 0.88)$, which yielded one point which has leverage measure  ten times greater than the average value when $n=40$,   two high leverage points when $n=80$ and three, $n=120$.  Here, we used as measure of leverage the leverage generalized \citep{Espinheira2008}.  Notice that in Scenarios 5, 6 and 7  the $P^2$ measure seems more able to identify correctly that the leverage points
affect the goodness of prediction than the  $P^2_{\beta\gamma}$ measure. On the order hand, the $P^2_{\beta\gamma}$ outperforms the $P^2$ in Scenario 8.  It is interesting to notice that in Scenario 5, which represents one covariate in both submodels, with  
the only one covariate of mean submodel had values generated from the $t_{(3)}$  occurs the smaller values of the three statistics. Thus, the statistics  correctly  lead  to the conclusion that as greatest  is the influence of leverage point in the data,  worst are the predictions and the model fit. 

\begin{table}
\caption{\label{T:T2} Statistic values. Model correctly specified. 
$g(\mu_t) =\log( {\mu_t}/{(1-\mu_t)}) \text{ and } \,h(\phi_t) =\log(\phi_t),$ $\,t=1,\ldots,n.$} 
{\tiny
\begin{tabular}{*{14}{|c|}}
&\text{Scenarios}&\multicolumn{3}{|c}{\text{Scenario 5}}&\multicolumn{3}{|c}{\text{Scenario 6}}&\multicolumn{3}{|c}{\text{Scenario 7}}&\multicolumn{3}{|c}{\text {Scenario 8}}\\\hline
&Mean&\multicolumn{3}{|c}{$g(\mu_t) = \beta_1 + \beta_2\,x_{t2} $}&\multicolumn{3}{|c}{$g(\mu_t) = \beta_1 + \beta_2\,x_{t2} $}&\multicolumn{3}{|c}{$g(\mu_t) = \beta_1 + \beta_2\,x_{t2} $}&\multicolumn{3}{|c}{$g(\mu_t) = \beta_1 + \beta_2\,x_{t2}+$}\\ 
&submodels&\multicolumn{3}{|c}{}&\multicolumn{3}{|c}{$+ \beta_3\,x_{t3}$}&\multicolumn{3}{|c}{$+ \beta_3\,x_{t3} + \beta_4\,x_{t4}$}&\multicolumn{3}{|c}{$\hskip-0.08truein\beta_3\,x_{t3} + \beta_4\,x_{t4}+ \beta_5\,x_{t5}\hskip-0.1truein$}\\ \hline 
&Dispersion&\multicolumn{3}{|c}{$h(\phi_t) = \gamma_1 + \gamma_2\,z_{t2}$}&\multicolumn{3}{|c}{$h(\phi_t) = \gamma_1 + \gamma_2\,z_{t2}$}&\multicolumn{3}{|c}{$h(\phi_t) = \gamma_1 + \gamma_2\,z_{t2}$}&\multicolumn{3}{|c}{$h(\phi_t) = \gamma_1 + \gamma_2\,z_{t2}+$}\\ 
&submodels&\multicolumn{3}{|c}{}&\multicolumn{3}{|c}{$+ \gamma_3\,z_{t3}$}&\multicolumn{3}{|c}{$+ \gamma_3\,z_{t3} + \gamma_4\,z_{t4}   $}&\multicolumn{3}{|c}{$\hskip-0.1truein\gamma_3\,z_{t3} + \gamma_4\,z_{t4} + \gamma_5\,z_{t5}\hskip-0.1truein$}\\\hline                                                              
$\mu$& \multicolumn{13}{|c}{$\mu \in(0.20, 0.88)$}\\ \hline
$n$&$\lambda$&20&50&100&20&50&100&20&50&100&20&50&100\\ \hline 
40 &$P^2$        &0.794&0.764&0.742&0.743&0.721&0.699&0.792&0.769&0.749&0.731&0.731&0.725\\
  &$P^2_{\beta\gamma}$      &0.823&0.812&0.806&0.772&0.762&0.755&0.850&0.843&0.838&0.819&0.824&0.826\\
  &$R^2_{LR}$&0.834&0.837&0.842&0.771&0.784&0.797&0.779&0.779&0.785&0.712&0.738&0.761\\\hline
80 &$P^2$        &0.773&0.739&0.714&0.702&0.674&0.649&0.745&0.715&0.687&0.646&0.642&0.630\\
  &$P^2_{\beta\gamma}$      &0.803&0.789&0.781&0.732&0.717&0.708&0.814&0.802&0.794&0.758&0.762&0.763\\
  &$R^2_{LR}$&0.840&0.844&0.850&0.783&0.796&0.810&0.789&0.790&0.796&0.724&0.749&0.772\\\hline
120&$P^2$        &0.766&0.731&0.704&0.688&0.657&0.630&0.729&0.696&0.665&0.615&0.609&0.596\\
  &$P^2_{\beta\gamma}$      &0.796&0.781&0.771&0.717&0.701&0.690&0.801&0.788&0.778&0.737&0.740&0.740\\
  &$R^2_{LR}$&0.842&0.846&0.852&0.786&0.799&0.813&0.793&0.793&0.799&0.727&0.753&0.775\\\hline
$\mu$      & \multicolumn{13}{|c}{$\mu \in (0.90, 0.99) $}\\ \hline
$n$&$\lambda$&20&50&100&20&50&100&20&50&100&20&50&100\\ \hline 
40 &$P^2$        &0.448&0.569&0.633&0.527&0.594&0.650&0.576&0.671&0.730&0.789&0.818&0.841\\   
  &$P^2_{\beta\gamma}$      &0.775&0.870&0.905&0.829&0.878&0.909&0.839&0.901&0.933&0.950&0.960&0.968\\
  &$R^2_{LR}$&0.455&0.557&0.617&0.432&0.494&0.557&0.353&0.445&0.513&0.454&0.501&0.544\\\hline
80 &$P^2$        &0.410&0.534&0.599&0.461&0.534&0.592&0.482&0.592&0.661&0.707&0.743&0.774\\
  &$P^2_{\beta\gamma}$      &0.770&0.862&0.898&0.809&0.861&0.895&0.804&0.879&0.915&0.928&0.942&0.954\\
  &$R^2_{LR}$&0.491&0.588&0.644&0.471&0.530&0.589&0.399&0.485&0.551&0.493&0.536&0.576\\\hline
120&$P^2$        &0.396&0.522&0.587&0.436&0.511&0.571&0.451&0.566&0.637&0.678&0.714&0.750\\
  &$P^2_{\beta\gamma}$      &0.767&0.859&0.895&0.801&0.855&0.890&0.794&0.872&0.909&0.921&0.935&0.948\\
  &$R^2_{LR}$&0.501&0.597&0.653&0.482&0.541&0.599&0.412&0.497&0.563&0.504&0.544&0.586\\\hline
$\mu$      & \multicolumn{13}{|c}{$\mu \in(0.005, 0.12)$}\\ \hline
$n$&$\lambda$&20&50&100&20&50&100&20&50&100&20&50&100\\ \hline 
40 &$P^2$        &0.680&0.769&0.811&0.641&0.692&0.732&0.647&0.739&0.797&0.800&0.832&0.862\\      
  &$P^2_{\beta\gamma}$      &0.218&0.298&0.367&0.257&0.296&0.341&0.281&0.332&0.387&0.409&0.442&0.486\\
  &$R^2_{LR}$&0.719&0.781&0.822&0.609&0.639&0.677&0.464&0.547&0.621&0.532&0.585&0.644\\\hline
80 &$P^2$        &0.657&0.748&0.792&0.584&0.639&0.683&0.567&0.675&0.742&0.721&0.763&0.804\\
  &$P^2_{\beta\gamma}$      &0.166&0.248&0.321&0.175&0.216&0.262&0.169&0.220&0.278&0.271&0.305&0.355\\
  &$R^2_{LR}$&0.743&0.754  & 0.761   &0.639&0.667&0.704&0.504&0.585&0.657&0.566&0.617&0.676\\\hline
120&$P^2$        &0.650&0.741&0.784&0.565&0.619&0.664&0.540&0.653&0.722&0.691&0.737&0.781\\
  &$P^2_{\beta\gamma}$      &0.150&0.232&0.306&0.148&0.189&0.236&0.132&0.183&0.242&0.225&0.260&0.311\\
  &$R^2_{LR}$&0.750& 0.760  & 0.774   &0.649&0.675&0.711&0.515&0.596&0.668&0.577&0.628& 0.689\\\hline 
     & \multicolumn{13}{|c}{\text {covariate values generated from }$t_{(3)}$  \quad $\mu \in(0.20. 0.88)$. }\\ \hline
$n$&$\lambda$&20&50&100&20&50&100&20&50&100&20&50&100\\ \hline 
40 &$P^2$        &0.426&0.401&0.385&0.733&0.705&0.680&0.624&0.603&0.585&0.775&0.773&0.768\\
  &$P^2_{\beta\gamma}$      &0.526&0.544&0.565&0.822&0.812&0.806&0.685&0.700&0.716&0.751&0.762&0.768\\
  &$R^2_{LR}$&0.515&0.555&0.593&0.756&0.772&0.787&0.641&0.671&0.697&0.741&0.776&0.800\\\hline
80 &$P^2$        &0.400&0.364&0.340&0.696&0.658&0.628&0.553&0.516&0.490&0.710&0.701&0.692\\
  &$P^2_{\beta\gamma}$      &0.633&0.618&0.613&0.793&0.779&0.769&0.750&0.744&0.740&0.673&0.680&0.686\\
  &$R^2_{LR}$&0.537&0.577&0.616&0.767&0.782&0.799&0.657&0.685&0.711&0.754&0.789&0.815\\\hline
120&$P^2$        &0.386&0.348&0.322&0.682&0.641&0.608&0.523&0.482&0.453&0.687&0.675&0.663\\
  &$P^2_{\beta\gamma}$      &0.639&0.622&0.614&0.783&0.767&0.756&0.742&0.732&0.726&0.644&0.650&0.655\\
  &$R^2_{LR}$&0.545&0.584&0.623&0.770&0.785&0.802&0.661&0.689&0.715&0.759&0.793&0.819\\\hline
\end{tabular}}
\end{table}

\begin{table}
\caption{\label{T:T3} Statistic values. True models:
$g(\mu_t) =\log( {\mu_t}/{(1-\mu_t)}) = \beta_1 + \beta_i\,x_{ti},\,\log(\phi_t) = \gamma_1 + \gamma_i\,z_{ti},$ $\,\,i=2,3,4,5,$ and $\,t=1,\ldots,n.$ Misspecified models: $\phi$ fixed .}
{\tiny
\begin{tabular}{*{14}{|c|}} 
&\text{Scenarios}&\multicolumn{3}{|c}{\text{Scenario 5}}&\multicolumn{3}{|c}{\text{Scenario 6}}&\multicolumn{3}{|c}{\text{Scenario 7}}&\multicolumn{3}{|c}{\text {Scenario 8}}\\\hline
&      &\multicolumn{3}{|c}{$g(\mu_t) = \beta_1 + \beta_2\,x_{t2}$}&\multicolumn{3}{|c}{$g(\mu_t) = \beta_1 + \beta_2\,x_{t2}$}&\multicolumn{3}{|c}{$g(\mu_t) = \beta_1 + \beta_2\,x_{t2}$}&\multicolumn{3}{|c}{$g(\mu_t) = \beta_1 + \beta_2\,x_{t2} +$}\\ 
&True&\multicolumn{3}{|c}{}&\multicolumn{3}{|c}{$+ \beta_3\,x_{t3}$}&\multicolumn{3}{|c}{$\beta_3\,x_{t3} + \beta_4\,x_{t4}$}&\multicolumn{3}{|c}{$\beta_3\,x_{t3} + \beta_4\,x_{t4} + \beta_5\,x_{t5}$}\\                                                             
&models&\multicolumn{3}{|c}{$h(\phi_t) = \gamma_1 + \gamma_2\,z_{t2}$}&\multicolumn{3}{|c}{$h(\phi_t) = \gamma_1 + \gamma_2\,z_{t2}$}&\multicolumn{3}{|c}{$h(\phi_t) = \gamma_1 + \gamma_2\,z_{t2}$}&\multicolumn{3}{|c}{$h(\phi_t) = \gamma_1 + \gamma_2\,z_{t2}+$}\\ 
& &\multicolumn{3}{|c}{}&\multicolumn{3}{|c}{$+ \gamma_3\,z_{t3}$}&\multicolumn{3}{|c}{$+ \gamma_3\,z_{t3} + \gamma_4\,z_{t4}$}&\multicolumn{3}{|c}{$\gamma_3\,z_{t3} + \gamma_4\,z_{t4} + \gamma_5\,z_{t5} $}\\\hline                                                                                                                                 
&Estimated&\multicolumn{3}{|c}{$g(\mu_t) = \beta_1 + \beta_2\,x_{t2}$}&\multicolumn{3}{|c}{$g(\mu_t) = \beta_1 + \beta_2\,x_{t2}$}&\multicolumn{3}{|c}{$g(\mu_t) = \beta_1 + \beta_2\,x_{t2}$}&\multicolumn{3}{|c}{$g(\mu_t) = \beta_1 + \beta_2\,x_{t2} +$}\\ 
&models&\multicolumn{3}{|c}{}&\multicolumn{3}{|c}{$+ \beta_3\,x_{t3}$}&\multicolumn{3}{|c}{$+ \beta_3\,x_{t3} + \beta_4\,x_{t4}$}&\multicolumn{3}{|c}{$\beta_3\,x_{t3} + \beta_4\,x_{t4}+ \beta_5\,x_{t5}$}\\  \hline                                                           
$\mu$      & \multicolumn{13}{|c}{$\mu \in(0.20, 0.88)$}\\ \hline
$n$&$\lambda$&20&50&100&20&50&100&20&50&100&20&50&100\\ \hline 
40 &$P^2$        &0.778&0.734&0.699&0.761&0.721&0.677&0.707&0.674&0.639&0.718&0.707&0.685\\
  &$P^2_{\beta\gamma}$      &0.792&0.761&0.740&0.752&0.717&0.684&0.775&0.757&0.735&0.776&0.768&0.752\\
  &$R^2_{LR}$&0.777&0.734&0.701&0.722&0.676&0.624&0.607&0.555&0.505&0.571&0.549&0.512\\\hline
80 &$P^2$        &0.759&0.711&0.671&0.728&0.682&0.630&0.650&0.608&0.562&0.643&0.626&0.594\\
  &$P^2_{\beta\gamma}$      &0.772&0.738&0.714&0.714&0.673&0.631&0.728&0.707&0.679&0.717&0.703&0.680\\
  &$R^2_{LR}$&0.781&0.739&0.707&0.732&0.687&0.637&0.630&0.582&0.533&0.600&0.579&0.544\\\hline
120&$P^2$        &0.753&0.703&0.660&0.717&0.669&0.614&0.631&0.584&0.534&0.617&0.598&0.560\\
  &$P^2_{\beta\gamma}$      &0.764&0.729&0.702&0.702&0.656&0.611&0.712&0.688&0.660&0.696&0.680&0.651\\
  &$R^2_{LR}$&0.783&0.741&0.708&0.735&0.690&0.640&0.637&0.589&0.541&0.608&0.588&0.552\\\hline
$\mu$      & \multicolumn{13}{|c}{$\mu \in (0.90, 0.99) $}\\ \hline
$n$&$\lambda$&20&50&100&20&50&100&20&50&100&20&50&100\\ \hline 
40 &$P^2$        &0.141&0.163&0.175&0.194&0.208&0.220&0.280&0.292&0.300&0.347&0.350&0.351\\
  &$P^2_{\beta\gamma}$      &0.274&0.349&0.390&0.314&0.360&0.395&0.433&0.460&0.480&0.560&0.550&0.545\\
  &$R^2_{LR}$&0.250&0.244&0.233&0.177&0.153&0.134&0.058&0.039&0.023&0.086&0.064&0.041\\\hline
80 &$P^2$        &0.093&0.114&0.127&0.115&0.127&0.136&0.165&0.172&0.176&0.215&0.211&0.209\\
  &$P^2_{\beta\gamma}$      &0.242&0.320&0.364&0.253&0.296&0.327&0.332&0.352&0.368&0.464&0.442&0.431\\
  &$R^2_{LR}$&0.275&0.268&0.257&0.213&0.191&0.171&0.115&0.097&0.082&0.162&0.139&0.118\\\hline
120&$P^2$        &0.077&0.098&0.111&0.089&0.100&0.109&0.125&0.130&0.133&0.170&0.163&0.159\\
  &$P^2_{\beta\gamma}$      &0.231&0.311&0.356&0.231&0.274&0.304&0.295&0.313&0.325&0.428&0.400&0.385\\
  &$R^2_{LR}$&0.282&0.275&0.265&0.225&0.203&0.182&0.131&0.114&0.098&0.181&0.157&0.138\\\hline
$\mu$      & \multicolumn{13}{|c}{$\mu \in(0.005, 0.12)$}\\ \hline
$n$&$\lambda$&20&50&100&20&50&100&20&50&100&20&50&100\\ \hline 
40 &$P^2$        &0.295&0.312&0.316&0.270&0.285&0.295&0.317&0.331&0.338&0.371&0.374&0.377\\
  &$P^2_{\beta\gamma}$      &0.119&0.123&0.128&0.168&0.172&0.179&0.243&0.252&0.257&0.278&0.292&0.301\\
  &$R^2_{LR}$&0.560&0.522&0.484&0.400&0.363&0.326&0.159&0.143&0.130&0.165&0.146&0.124\\\hline
80 &$P^2$        &0.272&0.288&0.290&0.202&0.217&0.226&0.205&0.214&0.218&0.246&0.239&0.234\\
  &$P^2_{\beta\gamma}$      &0.068&0.073&0.077&0.086&0.090&0.096&0.127&0.133&0.137&0.138&0.149&0.155\\
  &$R^2_{LR}$&0.603&0.576&0.545&0.439&0.422&0.399&0.216&0.207&0.201&0.245&0.223&0.204\\\hline
120&$P^2$        &0.264&0.280&0.283&0.177&0.194&0.203&0.166&0.171&0.175&0.202&0.189&0.180\\
  &$P^2_{\beta\gamma}$      &0.052&0.058&0.062&0.059&0.063&0.069&0.087&0.092&0.095&0.093&0.101&0.105\\
  &$R^2_{LR}$&0.618&0.596&0.570&0.449&0.439&0.427&0.231&0.222&0.220&0.266&0.2410& 0.2520\\\hline
\end{tabular}}
\end{table}

Finally, were carried out Monte Carlo simulations  to assess the performance of statistics  when the  dispersion modelling is  neglected.  To that end, the true data generating process considers  varying dispersion   
but a fixed dispersion beta regression is estimated; see
Table~\ref{T:T3}. 
In this case we have misspecification. Thus, we hope that the statistics display smaller values in comparison with Table~\ref{T:T3}.  In this sense, when  $\mu \in (0.005, 0.12)$, it is noteworthy
that the $P^2_{\beta\gamma}$ statistic  outperforms the $P^2$ statistic  identifying more emphatically the misspecification.  For the other hand, when $\mu \approx 1$ is the  $P^2$ statistic that emphasizes the poor prediction power of the model when the varying dispersion  is neglected.  But, in fact, the statistics 
behavior slightly  change. For instance, consider  $\mu\in(0.20, 0.88),$   $n=120,$    $\lambda = (20, 50, 100)$ 
and Scenario 4 (three covariates in both submodels).  When the dispersion is correctly modelled; Table~\ref{T:T2}, 
$P^2_{\beta\gamma} = (0.737, 0.740, 0.740)$ and  when the varying dispersion  is neglected; Table~\ref{T:T3}, $P^2_{\beta\gamma} = (0.696,0.680,0.651)$, respectively.

\section{Application}
In what follows we shall present an application based on real data.
The application relates to the distribution of natural gas for home usage
(e.g., in water heaters, ovens  and stoves) in S\~ao Paulo, Brazil. 
Such a distribution is based on two factors: the simultaneity factor ($F$) and   the total nominal power  of appliances that use natural gas, computed power $Q_{max}$.
Using these factors one obtains an indicator of gas release in a given tubulation section, namely:
$Q_p = F \times Q_{max}$.  The simultaneity factor  assumes values in $(0,1)$, and can be interpreted as the probability of simultaneous appliances usage.
Thus, based on $F$ the company that supplies the gas  decides how much gas
to supply to a given residential unit.

The data were analysed by \cite{Zerbinatti2008}, obtained from  the Instituto de Pesquisas Tecnol\'ogicas (IPT) and the Companhia de G\'as de S\~ao Paulo (COMG\'AS).  The response variable (y) are the simultaneity factors of 42 valid measurements of sampled households, and the covariate is the computed power.  The simultaneity factors ranged
from 0.02 to 0.46, being the  median   equals 0.07. \cite{Zerbinatti2008} modeled such data and concluded
that the best performing model was  the beta regression model based on logit link and log of  computed power used as covariate.
However, the author shows that the beta regression model can underpredict the response. Thus, \cite{Espinheira2014} argue that it is important to have at disposal prediction intervals that can be used with beta regressions. To that end, the authors built and evaluated bootstrap-based prediction intervals for the response for the class of beta regression models.  They applied the approach to the data on simultaneity factor. However, a important step in this case was the selection of the model with the best predictive power. 
To reach this aim the authors used a  simplified version of PRESS statistic given by 
$PRESS ={ {\sum_{t=1}^{42}(y_t - \widehat y_{(t)})^2} / { 42 }}$ which selected the same model of
\cite{Zerbinatti2008}.
Here we aim at selecting the better  predictive model to the data on simultaneity factor using the $P^2$ and $P^2_{\beta\gamma}$ statistics.
We also consider the $R^2_{LR}$ as the measure of goodness-of-fit model.
Since that the response is the simultaneity factor and the covariate $X_{2}$ is the log of
computed power, we considered four candidate models. At the outset, we consider two beta regression model with fixed dispersion, the first one
using logit link function  for $\mu$ and the second one using  log-log link function.  Then, in the following  two models the dispersion is nonconstant, with the logit and log-log submodels for $\mu$ and log submodels for $\phi$.  Then statistic values are presented in Table~\ref{T:T4}.
Here, we consider that the predictive power of the model is better when the measures $P^2$ and $P^2_{\beta\gamma}$ are close to one.

\begin{figure}
\centering
\makebox{\includegraphics[width=5.0truein]{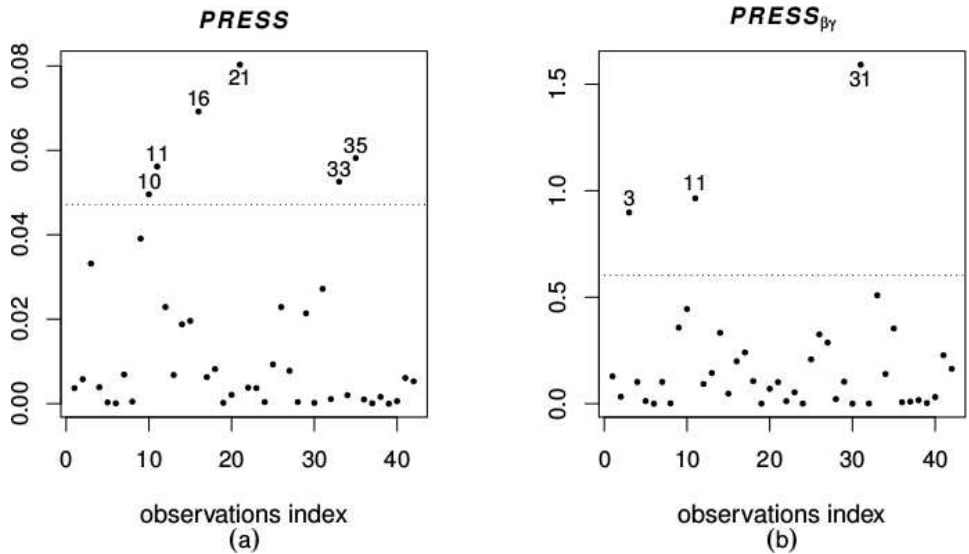}}
\caption{\label{fig:1} PRESS plots. PRESS (a) and PRESS$_{\beta\gamma}$ (b).}
\end{figure}

\begin{table}
\caption{\label{T:T4} Statistic values from the candidate models. Data on simultaneity factor} 
{\scriptsize
\begin{tabular}{*{5}{|c|}}
&\multicolumn{4}{|c}{\text{Candidate models}}\\\hline
Mean&{$\log( {\mu_t}/{(1-\mu_t)}) = $}&{$-\log(-\log {(\mu_t)}) = $}&{$\log( {\mu_t}/{(1-\mu_t)}) = $}&{$-\log(-\log {(\mu_t)}) = $}\\ submodel&{$\beta_1 + \beta_2\,x_{t2}$}&{$\beta_1 + \beta_2\,x_{t2}$}&{$\beta_1 + \beta_2\,x_{t2}$}&{$\beta_1 + \beta_2\,x_{t2}$}\\
Dispersion&&&{$\log(\phi_t) = $}&{$\log(\mu_t) = $}\\ 
submodel&&&{$\gamma_1 + \gamma_2\,x_{t2}$}&{$\gamma_1 + \gamma_2\,x_{t2}$}\\\hline
 $P^2$        &0.423&0.662 &0.461 &0.694  \\
 $P^2_{\beta\gamma}$      &0.100&0.100&0.131 &0.203  \\
 $R^2_{LR}$&0.701&0.683&0.701 &0.701  \\\hline
\end{tabular}}
\end{table}
The Table~\ref{T:T4} displays important informations.  First, we notice that by the $R^2_{RV}$ measures  three models equally fits well. 
Second, since that the responses are close to of lower limit of the standard unit interval
the statistics  display small values, in special the  $P^2_{\beta\gamma}$ statistic. Third, the $P^2$ and $P^2_{\beta\gamma}$ measures lead to the same conclusions,  selecting  the beta regression model with link log-log for the mean submodel and link log for the dispersion submodel,   as the best model to make prediction to the data on   simultaneity factor.
The maximum 
likelihood parameter estimates are   $\widehat{\beta}_1 = -0.63$, $\widehat{\beta}_2 = -0.31$, $\widehat{\gamma}_1 = 3.81$ and $\widehat{\gamma}_2 = 0.77$.
Furthermore, the estimative of  intensity of nonconstant dispersion  is $\widehat\lambda = 21.16$ (see (16)), such that $\widehat\phi_{\max} = 242.39$ and  $\widehat \phi_{\min} = 11.45$ .
Selected among the candidates  the best model in a predictive perspective, we still can use  the PRESS statistic to identifying which observations are more difficult to predict. 
In this sense, we plot the individual components of PRESS and PRESS$_{\beta\gamma}$ versus the observations index, Figure~\ref{fig:1}(a) and \ref{fig:1}(b), respectively.
Overall,  Figure~\ref{fig:1} shows that the cases 
3, 11 , 16, 21, 31, 33 and 35 arise as the observations with more predictive difficulty and are worthy of further investigation.

\section{Conclusion}
In this paper we develop the   $P^2$ and  $P^2 _ {\beta\gamma}$  based on two versions of PRESS 
statistics that we proposed for the class of beta regression models.   The $P^2$ coefficient consider 
the PRESS statistic  
based on ordinary residual from the Fisher's scoring iterative algorithm for estimating $\beta$
whereas $P^2 _ {\beta\gamma}$ is based on a new residual which is a combination of ordinaries residuals from the Fisher's scoring iterative algorithm for estimating $\beta$ and $\gamma$.
We have  presented the results of  Monte Carlo simulations carried out to evaluate the performance of predictive coefficients. Additionally, to access the goodness-of-fit model we used the $R^2_{LR}$. We consider different scenarios include misspecification of omitted covariates and  negligence of varying dispersion,  simultaneous increase in the number of covariates in the two submodels (mean and dispersion) and presence of leverage points in the data. 
Overall, the  coefficients $P^2$ and  $P^2 _ {\beta\gamma}$ perform similar and both showed  
enable to identify when the  model are not reliable or when is more difficult to make prediction. In this
situations, the  $R^2 _{LR}$ statistic also revels that the model does not fit well. 
It is noteworthy that when the response values are close to one or close to zero the  power predictive of the model is substantially affected even under correct specification. Finally,  an empirical application was
performed.
\section*{Acknowledgements}
We thank anonymous referees for comments and suggestions and gratefully acknowledge partial financial support from CNPq. 
\appendix
\section{Appendix A: Fisher's scoring iterative algorithm}
\label{apendice}
{\footnotesize
\noindent In what follows we shall present the score function and Fisher's information 
for $\beta$  and $\gamma$ in the class of varying dispersion beta regression models \citep{Ferrari2011}.  We shall also present results that are useful to the 
derivation of the residuals proposed in this paper. 
The log-likelihood function for model \eqref{eq1} is given by 
$\ell(\beta,\gamma) = \sum_{t=1}^n \ell_t(\mu_t,\phi_t),$
where
$\ell_t(\mu_t,\phi_t) = \log \Gamma(\phi_t) - \log
\Gamma(\mu_t \phi)
- \log \Gamma( (1-\mu_t) \phi_t) + (\mu_t\phi_t-1) \log y_t
+ \{(1-\mu_t)\phi_t - 1\} \log (1-y_t).$
The score function for $\beta$ is thus 
$U_{\beta}(\beta, \gamma) =  X^{\!\top}\Phi T (y^*-\mu^{*})$,
$X$ is an $n\times k$, $\Phi = {\rm diag}(\phi_1,\ldots,\phi_n)$ and $t$th elements of $y^*$ and $\mu^*$ being given in \eqref{eq5},
\begin{equation}
\label{A1}
T = {\rm diag}\{ 1/g'(\mu_1), \ldots, 1/g'(\mu_n)\};
\end{equation}
the score function for $\gamma$ can be written as 
$U_{\gamma}(\beta, \gamma) =  Z^{\!\top}Ha,$
where, $Z$ is an $n\times q$, $a_t$ being given in \eqref{eq7} and 
\begin{equation}
\label{A2}
H = {\rm diag}\{ 1/h'(\phi_1), \ldots, 1/h'(\phi_n)\}. 
\end{equation}
The components of Fisher's information matrix are
$K_{\beta\beta} = X^{\!\top}\Phi WX$, 
$K_{\beta\gamma}
= K_{\gamma\beta}^{\!\top} = X^{\!\top}CTHZ$ and $K_{\gamma\gamma} = Z^\top D Z$.
Here, $W = {\rm diag}\{ w_1, \ldots, w_n\}$; where
\begin{equation}
\label{A3}
w_t = \phi_t v_t [1 / \{g'(\mu_t)\}^2] \quad \text{and}\quad v_t = \left\{ \psi'(\mu_t\phi_t) + \psi'((1-\mu_t)\phi_t)\right\}.
\end{equation}
Also,  $C = {\rm diag}\{ c_1, \ldots, c_n\}$, with
$c_t = \phi_t \left\{ \psi'(\mu_t\phi_t)\mu_t - \psi'((1-\mu_t)\phi_t)
(1-\mu_t)\right\}$ and
$D={\rm diag}\{ d_1, \ldots, d_n\}$, with 
\begin{equation}
\label{A4}
 d_t = \varsigma_t \frac {1}{  \{h'(\mu_t)\}^2} \, \,\,\text{and}\,\, \,\varsigma_t = \left\{\psi'(\mu_t\phi_t)\mu_t^2 + \psi'((1-\mu_t)\phi_t)(1-\mu_t)^2
- \psi'(\phi_t)\right\}.
\end{equation}

The Fisher's scoring iterative schemes used for estimating 
$\beta$ and $\gamma$ can be written, respectively, as 
\begin{equation}
\label{A5}
\beta^{(m+1)} =\beta^{(m)} +  {(K_{\beta\beta}^{(m)})^{-1}}U_{\beta}^{(m)}(\beta) \quad \text{and}\quad \gamma^{(m+1)} =\gamma^{(m)} +  {(K_{\gamma\gamma}^{(m)})^{-1}}U_{\gamma}^{(m)}(\gamma),
\end{equation}
where $m=0,1,2,\ldots$ are the iterations that are performed until convergence, 
which occurs when the distance between $\beta^{(m+1)}$ and $\beta^{(m)}$ becomes 
smaller than a given small constant. 

It is important to note that the beta density \eqref{eq1} belongs to canonical two-parameter 
exponential family. Indeed, 
$f(y_t; \mu_t, \phi_t)= \exp\{\tau_1 T_1  + \tau_2 T_2-{\cal A}(\tau)\}(1 / y_t(1-y_t))$,
where $\tau = (\tau_1, \tau_2) = (\mu_t\phi_t, \phi_t)$,
$(T_1, T_2) = (\log\{Y_t / (1-Y_t)\},\log(1-Y_t))$   and
${\cal A}(\tau) =
\{ -\log \Gamma (\phi_t)+ \log \Gamma (\mu_t\phi_t) +  \log \Gamma ((1-\mu_t)\phi_t)\}$.
Thus, 
\begin{equation}
\label{A6}
{\rm E}(T_1) = {\rm E}(Y^*_t) ={\partial{\cal A}(\tau)/ \partial \tau_1} =  \psi(\mu_t\phi_t)
- \psi((1-\mu_t)\phi_t)= \mu_t^*,
\end{equation}

\begin{equation}
\label{A7}
{\rm E}(T_2) = {\rm E}(\log(1-Y_t)) = {\partial{\cal A}(\tau)/ \partial \tau_2} = \psi((1 - \mu_t)\phi_t) - \psi(\phi_t), 
\end{equation}
\begin{equation}
\label{A8}
{\rm Var}(T_1) = {\rm Var}(Y^*_t) ={\partial^2{\cal A}(\tau)/
\partial \tau_1^2}= \psi'(\mu_t\phi_t)
+ \psi'((1-\mu_t)\phi_t)= v_t,
\end{equation}
\begin{equation} 
\label{A9}
{\rm Var}(T_2) = {\rm Var}(\log(1-Y_t)) = {\partial^2{\cal A}(\tau)/
\partial \tau_1^2} = \psi'((1 - \mu_t)\phi_t) - \psi'(\phi_t), 
\end{equation}
and
\begin{equation}
\label{A10}
{\rm Cov}(T_1,T_2) = {\partial^2{\cal A}(\tau)/
\partial \tau_1 \partial \tau_2} = - \psi'((1 - \mu_t)\phi_t). 
\end{equation}
More details see \citet[p.\ 27]{lehmann1998}.}

%
\bibliographystyle{rss}
\bibliography{PRESS}

\begin{thebibliography}{15}
\expandafter\ifx\csname natexlab\endcsname\relax\def\natexlab#1{#1}\fi
\expandafter\ifx\csname url\endcsname\relax
  \def\url#1{\texttt{#1}}\fi
\expandafter\ifx\csname urlprefix\endcsname\relax\def\urlprefix{URL}\fi

\bibitem[{Akaike(1973)}]{Aka:1973}
Akaike, H. (1973) Information theory and an extension of the maximum likelihood
  principle.
\newblock In \textit{Second International Symposium on Information Theory
  (Tsahkadsor, 1971)}, 267--281. Budapest: Akad\'emiai Kiad\'o.

\bibitem[{Allen(1974)}]{All:1974}
Allen, D.~M. (1974) The relationship between variable selection and data
  augmentation and a method for prediction.
\newblock \textit{Technometrics}, \textbf{16}, 125--127.

\bibitem[{Bayer and Cribari-Neto(2014)}]{Bayer2014a}
Bayer, F.~M. and Cribari-Neto, F. (2014) {Model selection criteria in beta
  regression with varying dispersion}.
\newblock \urlprefix\url{http://arxiv.org/abs/1405.3718}.

\bibitem[{Cook and Weisberg(1982)}]{CookWeisberg82book}
Cook, R.~D. and Weisberg, S. (1982) \textit{Residuals and Influence in
  Regression}.
\newblock Chapman and Hall.

\bibitem[{Cribari-Neto and Zeileis(2010)}]{Cribari2010}
Cribari-Neto, F. and Zeileis, A. (2010) Beta regression in {R}.
\newblock \textit{Journal of Statistical Software}, \textbf{34}, 1--24.

\bibitem[{Espinheira et~al.(2008)Espinheira, Ferrari and
  Cribari-Neto}]{Espinheira2008}
Espinheira, P., Ferrari, S. and Cribari-Neto, F. (2008) On beta regression
  residuals.
\newblock \textit{Journal of Applied Statistics}, \textbf{35}, 407--419.

\bibitem[{Espinheira et~al.(2014)Espinheira, Ferrari and
  Cribari-Neto}]{Espinheira2014}
--- (2014) Bootstrap prediction intervals in beta regressions.
\newblock \textit{Computational Statistics}, \textbf{29}, 1263--1277.

\bibitem[{Ferrari and Cribari-Neto(2004)}]{Ferrari2004}
Ferrari, S. and Cribari-Neto, F. (2004) Beta regression for modelling rates and
  proportions.
\newblock \textit{Journal of Applied Statistics}, \textbf{31}, 799--815.

\bibitem[{Ferrari et~al.(2011)Ferrari, Espinheira and
  Cribari-Neto}]{Ferrari2011}
Ferrari, S. L.~P., Espinheira, P.~L. and Cribari-Neto, F. (2011) Diagnostic
  tools in beta regression with varying dispersion.
\newblock \textit{Statistica Neerlandica}, \textbf{65}, 337--351.

\bibitem[{Lehmann and Casella(1998)}]{lehmann1998}
Lehmann, E. and Casella, G. (1998) \textit{{Theory of Point Estimation}}.
\newblock Springer Verlag.

\bibitem[{Mediavilla et~al.(2008)Mediavilla, F and Shah}]{Mediavilla2008}
Mediavilla, F., F, L. and Shah, V.~A. (2008) A comparison of the coefficient of
  predictive power, the coefficient of determination and aic for linear
  regression.
\newblock In \textit{Decision Sciences Institute, Atlanta} (ed. K.~JE),
  1261--1266.

\bibitem[{Pregibon(1981)}]{pregibon1981}
Pregibon, D. (1981) Logistic regression diagnostics.
\newblock \textit{Ann. Statist.}, \textbf{9}, 705--724.

\bibitem[{Schwarz(1978)}]{Schwarz78}
Schwarz, G. (1978) {Estimating the dimension of a model}.
\newblock \textit{Annals of Statistics}, \textbf{6}, 461--464.

\bibitem[{Smithson and Verkuilen(2006)}]{Smithson2006}
Smithson, M. and Verkuilen, J. (2006) {A Better Lemon Squeezer?
  Maximum-Likelihood Regression With Beta-Distributed Dependent Variables}.
\newblock \textit{Psychological Methods}, \textbf{11}, 54--71.

\bibitem[{Zerbinatti(2008)}]{Zerbinatti2008}
Zerbinatti, L. (2008) \textit{Predi\c{c}\~{a}o de fator de simultaneidade
  atrav\'{e}s de modelos de regress\~{a}o para propor\c{c}\~{o}es
  cont\'{\i}nuas}.
\newblock Msc thesis, University of S\~{a}o Paulo.
\newblock
  \urlprefix\url{{http://www.teses.usp.br/teses/disponiveis/45/45133/}}.

\end{thebibliography}

\end{document}